\documentclass{article}

\usepackage{microtype}
\usepackage{graphicx}
\usepackage{subfigure}
\usepackage{booktabs} % for professional tables
\usepackage{stfloats}

\usepackage{float}
\usepackage{amsmath,amssymb,amsfonts}
\usepackage[utf8]{inputenc}

\usepackage{hyperref}

\usepackage[accepted]{icml2021}

\icmltitlerunning{Deep Random Projection Outlyingness for Unsupervised Anomaly Detection}

\begin{document}

\twocolumn[
\icmltitle{Deep Random Projection Outlyingness for Unsupervised Anomaly Detection}



\icmlsetsymbol{equal}{*}

\begin{icmlauthorlist}
\icmlauthor{Martin Bauw}{mines,thales}
\icmlauthor{Santiago Velasco-Forero}{mines}
\icmlauthor{Jesus Angulo}{mines}
\icmlauthor{Claude Adnet}{thales}
\icmlauthor{Olivier Airiau}{thales}
\end{icmlauthorlist}

\icmlaffiliation{mines}{Center
for Mathematical Morphology, MINES ParisTech, PSL Research University, France}
\icmlaffiliation{thales}{Thales LAS France, Advanced Radar Concepts, Limours, France}

\icmlcorrespondingauthor{Martin Bauw}{martin.bauw@mines-paristech.fr}

\icmlkeywords{anomaly detection, one-class classification, deep
learning, random projection}

\vskip 0.3in
]



\printAffiliationsAndNotice{\icmlEqualContribution} % otherwise use the standard text.

\begin{abstract}
Random projection is a common technique for designing algorithms in a variety of areas, including information retrieval, compressive sensing and measuring of outlyingness. In this work, the original random projection outlyingness measure is modified and associated with a neural network to obtain an unsupervised anomaly detection method able to handle multimodal normality. Theoretical and experimental arguments are presented to justify the choice of the anomaly score estimator. The performance of the proposed neural network approach is comparable to a state-of-the-art anomaly detection method. Experiments conducted on the MNIST, Fashion-MNIST and CIFAR-10 datasets show the relevance of the proposed approach.
\end{abstract}

% sentence deleted from PR version: "The contribution of adapted dropouts is investigated, along with the affine stability of the proposed method."

% + " and suggest a possible extension to a semi-supervised setup."

\section{Introduction}
\label{introduction}

% dimensionality reduction
When working with high-dimensional data, achieving relevant data analysis can translate into finding the right projections for the data. Discovering a good projection for a dataset amounts to revealing a helpful perspective for the task at hand. Defining what a good projection actually is and finding it can then be challenging. Common data processing tools such as principal component analysis (PCA) and linear discriminant analysis (LDA) are examples of approaches that look for effective projections to discriminate between samples. The appeal of projecting high-dimensional data to lower dimensional spaces reside in the fact that such projections constitute a way of avoiding the curse of dimensionality to a certain extent \cite{huber1985projection}. In the specific case of projecting data to a single dimension, the use of common one-dimensional statistics can also be a motivation. However, finding good projections can be costly and therefore leads to considering the possible contribution of random projections (RPs), since randomization is cheaper than optimization \cite{rahimi2009weighted}. 
%More specifically, RPs are computationally less expensive than PCA, and can become even cheaper if one considers sparse random projections \cite{bingham2001random}.

% classic uses of RP: dimensionality reduction, extreme learning machine
RPs have been commonly used for dimensionality reduction, typically in the context of compressed-sensing \cite{candes2006near}. In \cite{fowler2011anomaly}, random projections are used to conduct anomaly detection with the projected data representations. Using a great number of random projections, a stochastic approximation of the depth function introduced in \cite{donoho1992breakdown} can be obtained. A projection depth function associates a depth attribute to each data point available in a dataset, without explicitly estimating the underlying probability density function. Such depth directly translates into an ordering of the data points, from the most normal to the most outlying one. Theoretical results indicate a convergence between anomaly detection using a threshold on such a depth and the detection achieved with a Reed-Xioli (RX) anomaly detector \cite{velasco2012robust}, based on the Mahalanobis distance. This convergence supports the relevance of using an RP-based method for outlier detection. 
%The ordering induced by a projection depth function has notably been used to construct morphological operators, and applied to hyperspectral images \cite{velasco2012random}.

% ELN + training BN and only BN (RP in neural networks)
In addition, RP emerged as a way to define a new kind of neural network, extreme learning machines (ELM), where the parameters of a single hidden layer of neurons are frozen in their random initialization state. Those frozen parameters can be considered as RPs, and are said to enable fast learning and good generalization properties for the neural network \cite{huang2006extreme}. The power of random projections within neural networks is not limited to ELM and single layer neural networks. In \cite{wojcik2019training}, an initial RP layer is considered for high-dimensional real world datasets with various initialization schemes coming from the RP literature. Here again, the author's interest in RP stems from the possibility of creating relevant embeddings. The expressive power of random features in ResNet \cite{He_2016_CVPR} architectures has been demonstrated in \cite{frankle2020training}, while the authors focused on the performances obtained when training only the affine transformation included in the Batch Normalization (BN).
%[11] has demonstrated the expressive power of random features in ResNet [12] architectures, while focusing on the performances obtained when training only the affine transformation included in the Batch Normalization (BN). 
Intuitively, training only the affine transformation of the BN is equivalent to training the shifting and rescaling of random features. 
% In contrast with ELM, this setup uses a repetition of a succession of random features and trainable parameters.
% More general background literature on random weights in neural networks
Neural networks built with frozen random weights have already received much attention in the literature, with contributions like \cite{saxe2011random,giryes2016deep} trying to capture the actual contribution of random weights.

% Anomaly detection
Anomaly detection (AD) remains an active research field \cite{chandola2009anomaly, pimentel2014review, hendrycks2019deep,chalapathy2019deep, ruff2021unifying} and is closely related to novelty detection, out-of-distribution detection and noise removal. Each of those functions can provide a way to detect anomalies, depending on the task at hand and the nature of the anomalies. Anomalies are defined as unusual or atypical objects and events in \cite{goodfellow2016deep}, however it is important to note that various notions of anomalies and AD supervision exist \cite{chandola2009anomaly}. In this work, the unsupervised learning paradigm of \cite{deepSVDD} is chosen. The latter indicates that the absence of supervision implies the unavailability of labeled anomalies during training. The training data will therefore consist of normal samples, supposedly representative of the normality. To consider a more realistic scenario, the AD methods can be compared on polluted training sets, where anomalous samples will corrupt the hypothetically normal training data.

The rest of the paper is structured as follows: in section \ref{baseline}, the inspiration behind the proposed deep outlyingness and the state-of-the-art deep AD baseline are described. Section \ref{RPO-unsupAD} introduces the RP-based unsupervised outlyingness measure called deep random projection outlyingness (RPO).
% and develops several implementation modalities. 
Finally, section \ref{experiments} presents the results of our experiments which put forward the preference for a specific version of deep RPO.

\section{RPO and deep SVDD: baselines and inspirations}
\label{baseline}

% original RPO and deep SVDD

%\subsection{Random projection outlyingness}

Let us consider data points with a $d$-dimensional representation. A random projection can be used to bring the latter on a single dimension representation space. Multiple random projections can thus be used to obtain multiple representations of data points on a single dimension, allowing the computation of normalized distances to the dataset center point for the dataset samples on each projection. These normalized distances lead to the RPO proposed in \cite{donoho1992breakdown} and defined by \eqref{RPO_1}, where $x$ stands for a single data point, and $X$ is the data matrix gathering all data points.

\begin{figure}[ht]
\vskip 0.2in
\begin{center}
%\centerline{\includegraphics[width=\columnwidth]{rp_spirit.png}}
\centerline{\includegraphics[width=0.65\columnwidth]{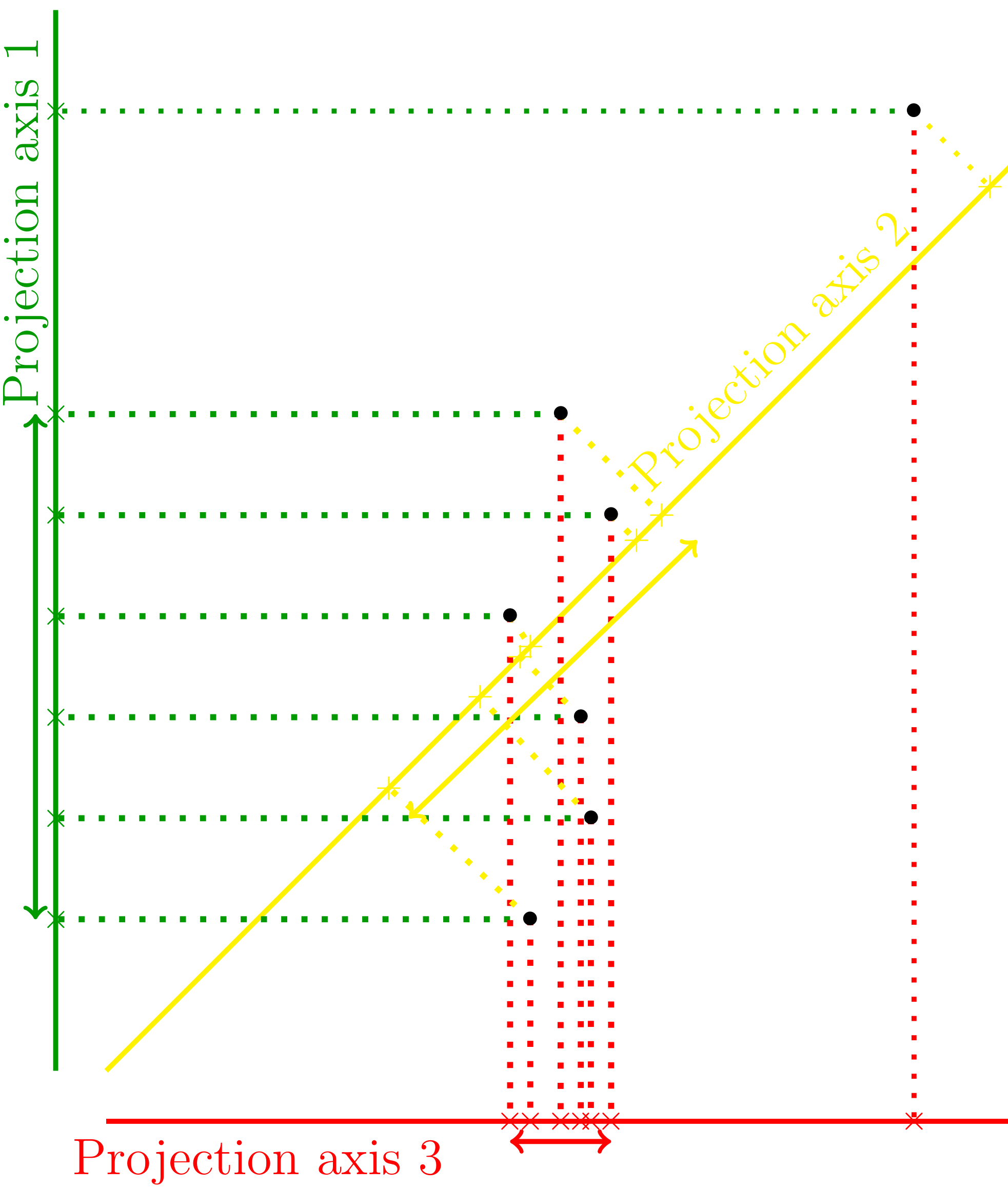}}
\caption{Illustration of the intuition behind the use of random projections. Once a set of 2D samples is projected, evaluating the normalized distance to the location estimator of each projection easily allows to detect the obvious outlier, the latter being positioned at greater distance from the location estimator on at least one random projection. One random projection is enough to raise the maximum seen in \eqref{RPO_1}. This depicts that multivariate outlyingness can translate into multiple univariate outlyingnesses.}
\label{rp-intuition}
\end{center}
\vskip -0.2in
\end{figure}

\begin{equation}
\label{RPO_1}
O(x;p, X) = \underset{u \in \mathbb{U}}{max} \dfrac{\vert u^Tx - MED(u^TX) \vert}{MAD(u^TX)}
\end{equation}

This quantity is obtained using a set of $p$ random projections $u$ of dimensionality $d$, written $\mathbb{U}$. All projection vectors $u$ have a unit norm, \emph{i.e.} $u \in \mathbb{S}^{d-1}$ with $\mathbb{S}^{d-1}=\{x\in\mathbb{R}^d:\vert\vert x \vert\vert_2 =1\}$. The maximum returns the greatest distance among the random projections to a location estimator normalized by a dispersion estimator, for robustness purposes here the median and the median absolute deviation (MAD). This can be interpreted as the worst-case distance to a robust location estimator. The word middle directly relates to the origin of \eqref{RPO_1} which stems from the definition of a statistical depth, the random projection depth, defined in \eqref{RPD}. The RP depth, as well as the RP outlyingness, establishes a center-outward ordering in a dataset:

\begin{equation}
\label{RPD}
RPD(x; X) = \dfrac{1}{1 + O(x; X)}
\end{equation}
with $O(x; X)$ the outlyingness \cite{rpd2} defined as
\begin{equation}
\label{RPO_0}
O(x; X) =  \underset{u \in \mathbb{S}^{d-1}}{sup} \dfrac{\vert u^Tx - MED(u^TX) \vert}{MAD(u^TX)}
\end{equation}

Originally, the random projection outlyingness is defined using a $sup$, but in practice it should be estimated with a maximum on  $p$ random projections on the $d-1$ dimensional sphere. The quantity $O(x;p, X)$ is therefore a stochastic approximation of $O(x; X)$ based on a finite set of random projections. The intuition of a naive RPO is depicted on Fig. \ref{rp-intuition}. The fundamental argument behind the interest for random projections to work on high-dimensional data is the Johnson-Lindenstrauss lemma \cite{william1984extensions}, 
%shown in \eqref{jllemma}, 
which guarantees the relative stability of the distance separating two data points between the input data space and the projected latent representations. 
%The lemma states, for a RP matrix $R \in \mathbb{R}^{k \times d}$ incorporating $k$ individual RPs, and two $d$-dimensional data points $x_1$ and $x_2$:
%
%\begin{equation}
%\label{jllemma}
%(1-\epsilon)\lVert x_1-x_2 \lVert^2_2 \leq \lVert Rx_1-Rx_2 \lVert^2_2 \leq (1+\epsilon)\lVert x_1-x_2 \lVert^2_2
%\end{equation}
%
%The factor $\epsilon$ depends on the number of random projections and the dimensionality of the projected space. 
% Said differently, the lemma ensures the Euclidean distance stability through random projections. 
%This implies a relative conservation of the data distribution in the projected and lower-dimensional space. Compared to other usual dimensionality reduction techniques, RPs are completely linear by the use of matrix $R$: auto-encoders harness non-linear activation functions, and principal components analysis discards components after linear transformations.

%\subsection{Deep support vector data description}
%\label{RPdim_dropout}

Deep support vector data description \cite{deepSVDD} is an unsupervised deep anomaly detection method. It is close to the original support vector data description (SVDD) \cite{svdd} and uses a neural network to project the supposedly normal training samples in a latent space so that all samples are within a normality hypersphere. The hypersphere is made as small as possible thanks to a suitable training loss, corresponding to

\begin{equation}
\label{DeepSVDD}
\begin{split}
\min_{W} \Bigg[ \dfrac{1}{n} \sum_{i=1}^n \vert \vert \Phi(x_i;W) - c \vert \vert^2 + \dfrac{\lambda}{2} \sum_{l=1}^L \vert \vert W^l \vert \vert^2_{F} \Bigg].
\end{split}
\end{equation}

In this equation, $c$ is the normality hypersphere center, determined by the mean latent coordinates of an initial forward pass of the training data. $\Phi$ represents the encoding neural network, $W$ its weights, $x_i$ the sample being projected, $l$ a layer index. A common regularization is performed on the network weights using the Frobenius norm, and is balanced through $\lambda$ with the main training objective. As put forward in \cite{deepSVDD}, which brought deep SVDD to the state of the art, several conditions need to be fulfilled in order to avoid the collapse of the normality hypersphere during training.
%\cite{deepSVDD}, which brought deep SVDD to the state of the art, puts forward several conditions that need to be fulfilled in order to avoid the collapse of the normality hypersphere during training. 
%A trivial example of such collapse consists in a neural network resulting in a constant latent representation, which is indeed resulting in a minimal training loss solution where the mean distance to the center is necessarily zero.

\section{From RPO to deep RPO}
\label{RPO-unsupAD}

% explain RPO ---> deep RPO (do not forget to explain RPO and deep SVDD baselines at the same time)

%\subsection{Deep RPO}

Whereas deep SVDD uses an Euclidean distance to the normality hypersphere center in the latent space, our proposition is to evaluate the distance to various location estimators provided by a diversity of non-trainable random projections. The outlyingness of \eqref{RPO_1} replaces the distance to a single hypersphere center to quantify abnormality in the latent space. The deep SVDD training objective of \eqref{DeepSVDD} becomes:

\begin{equation}
\label{deepRPO}
\begin{split}
\min_{W} 
\Bigg[ 
\dfrac{1}{n} \sum_{i=1}^n \left(\underset{u \in \mathbb{U}}{mean} \dfrac{\vert u^T \Phi(x_i;W) - MED(u^T\Phi(X;W)) \vert}{MAD(u^T\Phi(X;W))} \right)\\+ \frac{\lambda}{2} \sum_{l=1}^L \vert \vert W^l \vert \vert^2_{F} 
\Bigg]
\end{split}
\end{equation}

\noindent where both integrator operators \textit{max} from the original RPO and \textit{mean} in our proposition are studied in the conducted experiments. The estimator modification from $maximum$ to $mean$ can be interpreted as follows: the worst-case outlyingness over all the random projections considered, provided by the $max$, does not translate into an optimal learning objective, as it will be observed in the experiments. On the other hand, the $mean$ estimator can be seen as a way to take into account the abnormality indication of each random projection for every sample. In such a case, every perspective of the data points counts in the training loss, for every data point.

\section{Experimental setup and results}
\label{experiments}

Experiments were conducted on three common datasets in the machine learning community: MNIST, Fashion-MNIST and CIFAR-10. All experiments were conducted using PyTorch, on either of the following hardware configurations: AMD Ryzen 7 2700X with Nvidia RTX 2080, or Intel Xeon E5-2640 with Nvidia GTX Titan X. Table \ref{main-table} reports the main results of this work. RPO stands for the original RPO, described in \eqref{RPO_1} with its location estimator and spread measure, respectively the median and MAD, defined on the training dataset completely made of normal samples. This means RPO is adapted to a machine learning data paradigm, whereas the original RPO was meant to directly be applied to a test set in which there would not be a significant proportion of anomalies. The direct application of RPO to our test sets without determining the medians and MADs on the training data leads to performances next to randomness. Such unsupervised and untrained RPOs are therefore not represented in the results tables. This poor performance is due both to the inadequate balance between samples considered as anomalies and the normal ones, and the potentially insufficient number of RPs with respect to the input space. Indeed, the more the input space to which RPs are applied to is of high dimensionality, the more RPs you need to obtain an informative projected estimator \cite{gueguen2014local}. Most of the failure can however be attributed to the data balance of the test sets in this case. RPO is implemented using 1000 RPs.

%Apart from the results dedicated to the study of the influence of the number of RPs used in the latent RPO for deep RPO in table \ref{nbrRPs-table} and table \ref{nbrRPsCIFAR10-table}, RPO is implemented using 1000 RPs.

In table \ref{main-table}, RPO-max is the closest AD to the original RPO but as previously stated it is beforehand adapted to take into account training data. RPO-mean is the shallow equivalent of the proposed method, deep RPO-mean, which adds an encoding neural network in front of RPO in the AD process. The same goes for RPO-max and deep RPO-max, which constitutes a more direct descendant of the original RPO. The random projections tensor is initialized by a random realization of a standard normal distribution. Random projections leading to a single projected dimension are normalized, so that they belong to the unit sphere in accordance with \eqref{RPO_0}. 
% Multidimensional RPs, however, are not normalized as it is not required to keep the relative stability provided by the Johnson-Lindenstrauss lemma.

The input dimensionality for the shallow methods RPO-max and RPO-mean in table \ref{main-table} is the dimensionality of the flattened input images, i.e. 784 for MNIST and Fashion-MNIST, and 3072 for CIFAR10. Deep SVDD and deep RPO encode the input images into latent representations of 32 and 128 dimensions, for MNIST, Fashion-MNIST and CIFAR10 respectively, before projecting using RPs when RPs are used. Hyperparameters were directly inspired by the ones used by deep SVDD authors since their method constitutes the baseline to which the proposed method is compared. In particular, the encoding networks architectures are the ones used for MNIST, Fashion-MNIST and CIFAR10 for the original deep SVDD \cite{deepSVDD} and deep SAD \cite{deepSAD}. The weight decay hyperparameter was kept at $10^{-6}$, even though for deep RPO it did not have a great impact in our experiments when compared with trials where the decay had been removed. 
%Again, this choice was made with the motivation of keeping deep RPO as close to the baseline as possible to allow a fair comparison.

The metric used to evaluate the AD methods is the average AUC over several seeds, associated with a standard deviation, as can be found in the AD literature. One should keep in mind that when the number of classes defining normality increases, the datasets classes balance change. Before training, a validation set, made of 10\% of the original training set, is created using scikit-learn common split function. For all deep experiments, the retained test AUC is the one associated with the best epoch observed for the validation AUC as was done in \cite{chong2020simple}. AUCs reach either a convergence plateau or a maximum before dropping in 50 epochs, this number of epochs was thus chosen for all the experiments. This represents a substantial difference with the experimental setup proposed in \cite{deepSVDD}, where models were trained with many more epochs, benefited from pre-training accomplished using an auto-encoder, and a tailored preprocessing. The comparison between deep SVDD and deep RPO remains fair in this work since the network architecture is shared, along with the training hyperparameters. 
% The batch size was set to 128 for the three datasets considered.
% Il faut cependant prendre garde à ne pas s'en contenter, ces données étant assez simples et donc potentiellement éloignées des difficultés de la problématique réellement visée. Des considérations sur le danger de trop s'appuyer sur MNIST sont par exemple proposées dans \cite{ruff2020rethinking}: la présence d'information à différentes échelles au sein des échantillons est faible dans MNIST, alors que celle-ci peut se révéler importante dans des données plus complexes. Ce type d'information est pourtant très important dans le comportement des modèles, notamment vis-à-vis de leur réaction à une corruption des données.
For each experiment, a new seed is set and a random pick of normal classes is performed. This means that, unlike many other papers in the literature, the nature of normality can change every time a new seed is adopted. This additional diversity behind the average AUCs presented explains the high standard deviations observed in the results. 

% Auxiliary experiments, i.e. those not included in table \ref{main-table}, mostly rely on two Fashion-MNIST setups, where the normality is defined by either one or three classes. This is done due to space constraints, although the choice remains an informed one: CIFAR-10 was not selected because the multimodal AD remains an excessively complex task for the methods considered as table \ref{main-table} points out, while MNIST does not carry multiscale structure, making it a less interesting example \cite{ruff2020rethinking}. 

%While the multimodal normality improves the realism of the cases examined, the complexity of the datasets remain low, which adds to the unrealistic perfect purity of the training datasets, here completely made of normal samples. More complex data and labelling errors being unavoidable in real-world applications, further experiments should be considered on more realistic data and on polluted unsupervised AD training sets. Such experiments with shallow RPO can already be found in \cite{bauw2020unsupervised} on high resolution range profiles generated by a coastal surveillance radar.

%\subsection{The integrator estimator choice: mean versus maximum}

	\begin{table*}[t]
	%\begin{table*}[bp!]
	\caption{The integrator estimator choice: mean versus maximum. RPO, deep RPO and deep SVDD test AUCs on MNIST, Fashion-MNIST and CIFAR10 for 1 to 4  modes considered as normal, for 30 seeds (truncated mean AUC $\pm$ std).}
	\label{main-table}
	%\vskip 0.01in
	\begin{center}
	\begin{small}
	\begin{sc}
	%\scalebox{1.0}{
	\resizebox{1.\textwidth}{!}{
	\begin{tabular}{c|cc|cccc}
	\toprule
	\# modes & RPO-max (1) & RPO-mean & Deep SVDD & Deep RPO-max & Deep RPO-mean (2) & (2)-(1) \\
	
	\midrule
	MNIST - 1 & 	     84.64 $\pm$6.73& 84.12 $\pm$6.74& 88.60 $\pm$4.62& 87.96 $\pm$5.31& \textbf{90.10} $\pm$4.10 & 5.46\\
	\hspace*{0.5in} 2 & 75.27 $\pm$8.68& 72.83 $\pm$9.42& 84.35 $\pm$6.57& 83.79 $\pm$6.97& \textbf{85.36} $\pm$6.48 & 10.09\\
	\hspace*{0.5in} 3 & 69.67 $\pm$9.65& 66.92 $\pm$10.25& 81.23 $\pm$6.76& 80.16 $\pm$7.12& \textbf{81.60} $\pm$7.00 & 11.93\\
	\hspace*{0.5in} 4 & 66.54 $\pm$9.20& 63.60 $\pm$10.31& \textbf{78.89} $\pm$6.56& 77.35 $\pm$6.92& 78.65 $\pm$7.05 & 12.11\\
	
	\midrule
	F-MNIST - 1 		& 89.19 $\pm$5.81& 89.73 $\pm$5.79& 90.45 $\pm$5.76& 90.17 $\pm$6.09& \textbf{91.13} $\pm$5.20& 1.94\\
	\hspace*{0.61in} 2 & 78.52 $\pm$8.39& 76.47 $\pm$8.38& 85.24 $\pm$6.45& 84.57 $\pm$7.01& \textbf{85.81} $\pm$6.36& 7.29\\
	\hspace*{0.61in} 3 & 71.06 $\pm$7.38& 69.37 $\pm$7.64& 80.30 $\pm$6.99& 80.64 $\pm$6.69& \textbf{81.28} $\pm$6.40& 10.22\\
	\hspace*{0.61in} 4 & 67.58 $\pm$5.89& 65.79 $\pm$6.55& 77.30 $\pm$4.99& 77.53 $\pm$5.07& \textbf{77.82} $\pm$5.34& 10.24\\
	
	\midrule
	CIFAR10 - 1 		& 57.62 $\pm$10.96& 58.62 $\pm$9.43& \textbf{64.15} $\pm$7.38& 60.22 $\pm$7.00& 63.14 $\pm$7.30& 5.52\\
	\hspace*{0.61in} 2 & 53.85 $\pm$9.49& 53.81 $\pm$7.61& 56.37 $\pm$9.25& 55.66 $\pm$8.54& \textbf{56.46} $\pm$8.89& 2.61\\
	\hspace*{0.61in} 3 & 52.20 $\pm$6.95& 52.53 $\pm$5.08& 54.16 $\pm$6.94& 53.87 $\pm$6.20& \textbf{54.30} $\pm$6.80& 2.10\\
	\hspace*{0.61in} 4 & 51.88 $\pm$5.91& 52.32 $\pm$4.97& 53.64 $\pm$5.97& 53.71 $\pm$5.78& \textbf{53.88} $\pm$5.89& 2.00\\
	
	\bottomrule	
	\end{tabular}
	}
	\end{sc}
	\end{small}
	\end{center}
	\vskip -0.1in
	\end{table*}
	
The results of experiments over the three datasets considered, with 30 seeds per experimental setup, are gathered in table \ref{main-table}. The latter demonstrates the superiority of the $mean$ over the $max$ as an estimator for RPO when working with the deep RPO setup. The shallow RPO setup, on the other hand, suggests better performances can be obtained using a $max$. The neural network thus favours a loss balanced over all the single projected outlyingnesses. Moreover, the increasing AUC gap between deep RPO and shallow RPO ADs for MNIST and Fashion-MNIST supports the hypothesis that the encoding neural network allows RPO to face multimodal normality in AD. The growing gap in the last column is not observed for CIFAR10, however this failure is likely to stem from the excessive difficulty of the AD task rather than from an inability of deep RPO. 
% Additional results are proposed in \ref{nbrRPsCIFAR10-table} to show that the poor performances obtained for CIFAR10 are not related to a lack of RPs for the latent RPO, even though a higher latent dimensionality is adopted for CIFAR10 without increasing the number of RPs. 
The better performance of deep RPO-mean compared to deep SVDD places the proposed method at the state-of-the-art level.

\section{Conclusion}
\label{conclusion}

An adaptation of the classic outlyingness score based on random projections is proposed. In order to adapt the outlyingness score and obtain anomaly detection performances similar to the state of the art, the estimator is modified and a neural network is trained to encode the data in a latent space of lower dimensionality where the random projections outlyingness is redefined. This work emphasizes the possibility of adapting simple abnormality measures to complex and realistic anomaly detection tasks in which normality is multimodal. The experiments conducted on MNIST, Fashion-MNIST and CIFAR10 show a light improvement in performance with respect to Deep SVDD and suggest that the task of anomaly detection in a fully unsupervised framework, in the case of multimodal normality, remains a challenge. The relative success of the proposed approach highlights the relevance of random projections and more generally of untrained transformations in neural networks, when they are associated with a well chosen trainable architecture.

\bibliography{mybibfile}

\begin{thebibliography}{28}
\providecommand{\natexlab}[1]{#1}
\providecommand{\url}[1]{\texttt{#1}}
\expandafter\ifx\csname urlstyle\endcsname\relax
  \providecommand{\doi}[1]{doi: #1}\else
  \providecommand{\doi}{doi: \begingroup \urlstyle{rm}\Url}\fi

\bibitem[Candes \& Tao(2006)Candes and Tao]{candes2006near}
Candes, E.~J. and Tao, T.
\newblock Near-optimal signal recovery from random projections: Universal
  encoding strategies?
\newblock \emph{IEEE transactions on information theory}, 52\penalty0
  (12):\penalty0 5406--5425, 2006.

\bibitem[Chalapathy \& Chawla(2019)Chalapathy and Chawla]{chalapathy2019deep}
Chalapathy, R. and Chawla, S.
\newblock Deep learning for anomaly detection: A survey.
\newblock \emph{arXiv preprint arXiv:1901.03407}, 2019.

\bibitem[Chandola et~al.(2009)Chandola, Banerjee, and
  Kumar]{chandola2009anomaly}
Chandola, V., Banerjee, A., and Kumar, V.
\newblock Anomaly detection: A survey.
\newblock \emph{ACM computing surveys (CSUR)}, 41\penalty0 (3):\penalty0 1--58,
  2009.

\bibitem[Chong et~al.(2020)Chong, Ruff, Kloft, and Binder]{chong2020simple}
Chong, P., Ruff, L., Kloft, M., and Binder, A.
\newblock Simple and effective prevention of mode collapse in deep one-class
  classification.
\newblock In \emph{2020 International Joint Conference on Neural Networks
  (IJCNN)}, pp.\  1--9. IEEE, 2020.

\bibitem[Donoho et~al.(1992)Donoho, Gasko, et~al.]{donoho1992breakdown}
Donoho, D.~L., Gasko, M., et~al.
\newblock Breakdown properties of location estimates based on halfspace depth
  and projected outlyingness.
\newblock \emph{The Annals of Statistics}, 20\penalty0 (4):\penalty0
  1803--1827, 1992.

\bibitem[Fowler \& Du(2011)Fowler and Du]{fowler2011anomaly}
Fowler, J.~E. and Du, Q.
\newblock Anomaly detection and reconstruction from random projections.
\newblock \emph{IEEE Transactions on Image Processing}, 21\penalty0
  (1):\penalty0 184--195, 2011.

\bibitem[Frankle et~al.(2020)Frankle, Schwab, and Morcos]{frankle2020training}
Frankle, J., Schwab, D.~J., and Morcos, A.~S.
\newblock Training batchnorm and only batchnorm: On the expressive power of
  random features in cnns.
\newblock \emph{arXiv preprint arXiv:2003.00152}, 2020.

\bibitem[Giryes et~al.(2016)Giryes, Sapiro, and Bronstein]{giryes2016deep}
Giryes, R., Sapiro, G., and Bronstein, A.~M.
\newblock Deep neural networks with random gaussian weights: A universal
  classification strategy?
\newblock \emph{IEEE Transactions on Signal Processing}, 64\penalty0
  (13):\penalty0 3444--3457, 2016.

\bibitem[Goodfellow et~al.(2016)Goodfellow, Bengio, and
  Courville]{goodfellow2016deep}
Goodfellow, I., Bengio, Y., and Courville, A.
\newblock \emph{Deep learning}.
\newblock MIT press, 2016.

\bibitem[Gueguen et~al.(2014)Gueguen, Velasco-Forero, and
  Soille]{gueguen2014local}
Gueguen, L., Velasco-Forero, S., and Soille, P.
\newblock Local mutual information for dissimilarity-based image segmentation.
\newblock \emph{Journal of mathematical imaging and vision}, 48\penalty0
  (3):\penalty0 625--644, 2014.

\bibitem[He et~al.(2016)He, Zhang, Ren, and Sun]{He_2016_CVPR}
He, K., Zhang, X., Ren, S., and Sun, J.
\newblock Deep residual learning for image recognition.
\newblock In \emph{The IEEE Conference on Computer Vision and Pattern
  Recognition (CVPR)}, June 2016.

\bibitem[Hendrycks et~al.(2019)Hendrycks, Mazeika, and
  Dietterich]{hendrycks2019deep}
Hendrycks, D., Mazeika, M., and Dietterich, T.
\newblock Deep anomaly detection with outlier exposure.
\newblock In \emph{International Conference on Learning Representations}, 2019.

\bibitem[Huang et~al.(2006)Huang, Zhu, and Siew]{huang2006extreme}
Huang, G.-B., Zhu, Q.-Y., and Siew, C.-K.
\newblock Extreme learning machine: theory and applications.
\newblock \emph{Neurocomputing}, 70\penalty0 (1-3):\penalty0 489--501, 2006.

\bibitem[Huber(1985)]{huber1985projection}
Huber, P.~J.
\newblock Projection pursuit.
\newblock \emph{The annals of Statistics}, pp.\  435--475, 1985.

\bibitem[Pimentel et~al.(2014)Pimentel, Clifton, Clifton, and
  Tarassenko]{pimentel2014review}
Pimentel, M.~A., Clifton, D.~A., Clifton, L., and Tarassenko, L.
\newblock A review of novelty detection.
\newblock \emph{Signal Processing}, 99:\penalty0 215--249, 2014.

\bibitem[Rahimi \& Recht(2009)Rahimi and Recht]{rahimi2009weighted}
Rahimi, A. and Recht, B.
\newblock Weighted sums of random kitchen sinks: Replacing minimization with
  randomization in learning.
\newblock In \emph{Advances in neural information processing systems}, pp.\
  1313--1320, 2009.

\bibitem[Ruff et~al.(2018)Ruff, Vandermeulen, Goernitz, Deecke, Siddiqui,
  Binder, M{\"u}ller, and Kloft]{deepSVDD}
Ruff, L., Vandermeulen, R., Goernitz, N., Deecke, L., Siddiqui, S.~A., Binder,
  A., M{\"u}ller, E., and Kloft, M.
\newblock Deep one-class classification.
\newblock In Dy, J. and Krause, A. (eds.), \emph{Proceedings of the 35th
  International Conference on Machine Learning}, volume~80 of \emph{Proceedings
  of Machine Learning Research}, pp.\  4393--4402. PMLR, 10--15 Jul 2018.
\newblock URL \url{http://proceedings.mlr.press/v80/ruff18a.html}.

\bibitem[Ruff et~al.(2020{\natexlab{a}})Ruff, Vandermeulen, Franks, M{\"u}ller,
  and Kloft]{ruff2020rethinking}
Ruff, L., Vandermeulen, R.~A., Franks, B.~J., M{\"u}ller, K.-R., and Kloft, M.
\newblock Rethinking assumptions in deep anomaly detection.
\newblock \emph{arXiv preprint arXiv:2006.00339}, 2020{\natexlab{a}}.

\bibitem[Ruff et~al.(2020{\natexlab{b}})Ruff, Vandermeulen, G{\"o}rnitz,
  Binder, M{\"u}ller, M{\"u}ller, and Kloft]{deepSAD}
Ruff, L., Vandermeulen, R.~A., G{\"o}rnitz, N., Binder, A., M{\"u}ller, E.,
  M{\"u}ller, K.-R., and Kloft, M.
\newblock Deep semi-supervised anomaly detection.
\newblock In \emph{International Conference on Learning Representations},
  2020{\natexlab{b}}.

\bibitem[{Ruff} et~al.(2021){Ruff}, {Kauffmann}, {Vandermeulen}, {Montavon},
  {Samek}, {Kloft}, {Dietterich}, and {Müller}]{ruff2021unifying}
{Ruff}, L., {Kauffmann}, J.~R., {Vandermeulen}, R.~A., {Montavon}, G., {Samek},
  W., {Kloft}, M., {Dietterich}, T.~G., and {Müller}, K.~R.
\newblock A unifying review of deep and shallow anomaly detection.
\newblock \emph{Proceedings of the IEEE}, pp.\  1--40, 2021.
\newblock \doi{10.1109/JPROC.2021.3052449}.

\bibitem[Saxe et~al.(2011)Saxe, Koh, Chen, Bhand, Suresh, and
  Ng]{saxe2011random}
Saxe, A.~M., Koh, P.~W., Chen, Z., Bhand, M., Suresh, B., and Ng, A.~Y.
\newblock On random weights and unsupervised feature learning.
\newblock In \emph{ICML}, volume~2, pp.\ ~6, 2011.

\bibitem[Tax \& Duin(2004)Tax and Duin]{svdd}
Tax, D.~M. and Duin, R.~P.
\newblock Support vector data description.
\newblock \emph{Machine learning}, 54\penalty0 (1):\penalty0 45--66, 2004.

\bibitem[Tran et~al.(2017)Tran, Mai, et~al.]{tran2017anomaly}
Tran, K.~P., Mai, A.~T., et~al.
\newblock Anomaly detection in wireless sensor networks via support vector data
  description with mahalanobis kernels and discriminative adjustment.
\newblock In \emph{2017 4th NAFOSTED Conference on Information and Computer
  Science}, pp.\  7--12. IEEE, 2017.

\bibitem[Van~Aelst \& Rousseeuw(2009)Van~Aelst and Rousseeuw]{van2009minimum}
Van~Aelst, S. and Rousseeuw, P.
\newblock Minimum volume ellipsoid.
\newblock \emph{Wiley Interdisciplinary Reviews: Computational Statistics},
  1\penalty0 (1):\penalty0 71--82, 2009.

\bibitem[Velasco-Forero \& Angulo(2012)Velasco-Forero and
  Angulo]{velasco2012robust}
Velasco-Forero, S. and Angulo, J.
\newblock Robust rx anomaly detector without covariance matrix estimation.
\newblock In \emph{2012 4th Workshop on Hyperspectral Image and Signal
  Processing: Evolution in Remote Sensing (WHISPERS)}, pp.\  1--4. IEEE, 2012.

\bibitem[William \& Lindenstrauss(1984)William and
  Lindenstrauss]{william1984extensions}
William, B.~J. and Lindenstrauss, J.
\newblock Extensions of lipschitz mapping into hilbert space.
\newblock \emph{Contemporary mathematics}, 26\penalty0 (189-206):\penalty0
  323--341, 1984.

\bibitem[W{\'o}jcik \& Kurdziel(2019)W{\'o}jcik and
  Kurdziel]{wojcik2019training}
W{\'o}jcik, P.~I. and Kurdziel, M.
\newblock Training neural networks on high-dimensional data using random
  projection.
\newblock \emph{Pattern Analysis and Applications}, 22\penalty0 (3):\penalty0
  1221--1231, 2019.

\bibitem[Zuo et~al.(2003)]{rpd2}
Zuo, Y. et~al.
\newblock Projection-based depth functions and associated medians.
\newblock \emph{The Annals of Statistics}, 31\penalty0 (5):\penalty0
  1460--1490, 2003.

\end{thebibliography}
\bibliographystyle{icml2021}

\clearpage

\appendix
\twocolumn[
\icmltitle{Appendix}
]
%\section{Do \emph{not} have an appendix here}
%
%\textbf{\emph{Do not put content after the references.}}
%%
%Put anything that you might normally include after the references in a separate
%supplementary file.
%
%We recommend that you build supplementary material in a separate document.
%If you must create one PDF and cut it up, please be careful to use a tool that
%doesn't alter the margins, and that doesn't aggressively rewrite the PDF file.
%pdftk usually works fine. 
%
%\textbf{Please do not use Apple's preview to cut off supplementary material.} In
%previous years it has altered margins, and created headaches at the camera-ready
%stage. 
%%%%%%%%%%%%%%%%%%%%%%%%%%%%%%%%%%%%%%%%%%%%%%%%%%%%%%%%%%%%%%%%%%%%%%%%%%%%%%%
%%%%%%%%%%%%%%%%%%%%%%%%%%%%%%%%%%%%%%%%%%%%%%%%%%%%%%%%%%%%%%%%%%%%%%%%%%%%%%%

Auxiliary experiments, i.e. those not included in table \ref{main-table}, mostly rely on two Fashion-MNIST setups, where the normality is defined by either one or three classes. CIFAR-10 was not selected because the multimodal AD remains an excessively complex task for the methods considered as table \ref{main-table} points out, while MNIST does not carry multiscale structure, making it a less interesting example \cite{ruff2020rethinking}. In this appendix, apart from the results dedicated to the study of the influence of the number of RPs used in the latent RPO for deep RPO in table \ref{nbrRPs-table} and table \ref{nbrRPsCIFAR10-table}, RPO is implemented using 1000 RPs.

\section{Random projections output dimensionality and quantity}

\subsection{Multidimensional random projections description}

In the case of random projections leading to a single output dimension, we have the following setting: if $d$ is the data samples dimensionality, and $m$ the random projection output dimensionality, a random projection $u$ with $m=1$ will lead to a projected coordinate $u^Tx$ for any individual sample $x$ with $d$ dimensions. This projected coordinate can then be compared to a location estimator computed with the application of $u$ on all the available samples, $u^Tx - MED(u^TX)$ forms an example where the location estimator chosen is the median. 

On the other hand, for multidimensional random projections, i.e. $m>1$, a covariance matrix $C$ can be harnessed to obtain a more subtle distance to location estimators. Each sample $x$ can then be associated with a robust Mahalanobis distance $\sqrt{(u^Tx - MED)^T C^{-1} (u^Tx - MED)}$. In this configuration, each sample has an outlyingness based on $m$ projected coordinates per RP. Each of the projected coordinates is compared to a location estimator determined on each random projection dimension. One projected location estimator is thus computed over all data samples, for a single output dimension of the random projections in use. This transforms the training objective \eqref{deepRPO}, the distance to the median in the numerator becoming:

\begin{equation}
\label{deepRPO_cov}
\sqrt{(u^T \Phi(x_i) - MED)^T \ C^{-1} \ (u^T \Phi(x_i) - MED)}
\end{equation}
%\sqrt{\vert u^T \Phi(x_i) - MED \vert^T \ C^{-1} \ \vert u^T \Phi(x_i) - MED \vert}

The training loss now has the possibility to incorporate multidimensional projected representations for data samples, enabling additional latent representation flexibility. One should note that whereas deep SVDD is built on top of a latent normality hypersphere, deep RPO harnesses a latent normality ellipsoid \cite{van2009minimum}, each of the latent dimensions being subject to specific localization and spread parameters. An SVDD adaptation where the latent distances are computed using a Mahalanobis distance has been proposed in \cite{tran2017anomaly}, but the latter does not encode data with a neural network. Combining the deep version of SVDD with a Mahalanobis score would be another way to achieve a trainable latent normality representation based on an ellipsoid.

\subsection{Experiments}

	\begin{table}[H]
	\caption{Deep RPO test AUCs with varying RP latent dimensionality for the two estimators studied on Fashion-MNIST for 30 seeds (truncated mean AUC $\pm$ std).}
	\label{dimRPs-table}
	%\vskip 0.01in
	\begin{center}
	\begin{small}
	\begin{sc}
	\scalebox{1.0}{
	\begin{tabular}{lccccc}
	\toprule
	\# modes & 1 & 3 \\
	
	\midrule
	MAX - 1D RPs	& 90.17 $\pm$6.09& 80.63 $\pm$6.68\\
	MAX - 2D RPs	& 89.40 $\pm$6.43& 79.63 $\pm$7.08\\
	MAX - 4D RPs	& 89.47 $\pm$6.45& 79.60 $\pm$7.06\\
	\midrule
	MEAN - 1D RPs	& \textbf{91.13} $\pm$5.20& \textbf{81.28} $\pm$6.40\\
	MEAN - 2D RPs	& 90.36 $\pm$5.79& 80.44 $\pm$6.65\\
	MEAN - 4D RPs	& 90.24 $\pm$5.86& 80.44 $\pm$6.60\\

	\bottomrule	
	\end{tabular}
	}
	\end{sc}
	\end{small}
	\end{center}
	\vskip -0.1in
	\end{table}
	
Table \ref{dimRPs-table} reminds the point of switching from a supremum to an average for the AD score estimator, while revealing that RPs with muldimensional projected representations do not yield test AUC increases. No interest was paid to random projections with a latent dimensionality higher than four in order to avoid considerations pertaining to the quality of the covariance matrix of \eqref{deepRPO_cov}.

	\begin{table}[H]
	\caption{Deep RPO-mean test AUCs with varying number of RPs for the latent space RPO on Fashion-MNIST for 20 seeds (truncated mean AUC $\pm$ std).}
	\label{nbrRPs-table}
	%\vskip 0.01in
	\begin{center}
	\begin{small}
	\begin{sc}
	\scalebox{1.0}{
	\begin{tabular}{lccccc}
	\toprule
	\# modes & 1 & 3 \\
	
	\midrule
	100 RPs		& 90.25 $\pm$5.18& 81.70 $\pm$6.73\\
	500 RPs		& \textbf{90.46} $\pm$5.21& \textbf{81.96} $\pm$6.67\\
	1000 RPs	& 90.30 $\pm$5.25& 81.67 $\pm$6.87\\
	2000 RPs	& 90.42 $\pm$5.19& 81.83 $\pm$6.83\\

	\bottomrule	
	\end{tabular}
	}
	\end{sc}
	\end{small}
	\end{center}
	\vskip -0.1in
	\end{table}
	
Results in table \ref{nbrRPs-table} indicate an adequate number of RPs was chosen to implement the RP outlyingness for the encoded data. A slight AUC increase has been achieved by decreasing the number of random projections shared by the rest of the experiments, i.e. 1000, possibly indicating the approximate minimum number of RPs necessary to handle the dimensionality of the neural network encoded latent space. One can think that using the minimum number of RPs to handle the RPO score input dimensionality in a deep setup constitutes a sensible strategy since it avoids superfluous parameters without hurting the outlyingness measure.

	\begin{table}[H]
	\caption{Deep RPO-mean on CIFAR10 for 30 seeds, with either 1000 or 3000 RPs for RPO (truncated mean AUC $\pm$ std). The data being more complex, more RPs were used to verify whether a simple increase in the number of RPs could lead to better performances, without success.}
	\label{nbrRPsCIFAR10-table}
	%\vskip 0.01in
	\begin{center}
	\begin{small}
	\begin{sc}
	\scalebox{1.0}{
	\begin{tabular}{lcc}
	\toprule
	\# modes & 1000 RPs & 3000 RPs \\

	\midrule
%	CIFAR10 - 1 		 & 63.14 $\pm$7.30& 63.18 $\pm$7.49\\
%	\hspace*{0.725in} 2 & 56.46 $\pm$8.89& 56.44 $\pm$8.92\\
%	\hspace*{0.725in} 3 & 54.30 $\pm$6.80& 54.32 $\pm$6.74\\
%	\hspace*{0.725in} 4 & 53.88 $\pm$5.89& 54.00 $\pm$5.98\\
	1 & 63.14 $\pm$7.30& 63.18 $\pm$7.49\\
	2 & 56.46 $\pm$8.89& 56.44 $\pm$8.92\\
	3 & 54.30 $\pm$6.80& 54.32 $\pm$6.74\\
	4 & 53.88 $\pm$5.89& 54.00 $\pm$5.98\\
	
	\bottomrule	
	\end{tabular}
	}
	\end{sc}
	\end{small}
	\end{center}
	\vskip -0.1in
	\end{table}
	
As announced, table \ref{nbrRPsCIFAR10-table} suggests that the poor performances observed on CIFAR10 do not stem from an insufficient number of random projections, eventhough the greater latent dimensionality used for this dataset encoding could be expected to call for additional model complexity. These results emphasize the difficulty of the learning task considered when it comes to more realistic multimodal data.

\section{Projections and components dropouts}

\subsection{Dropouts description}

Picking up the previously introduced notation regarding random projections, $d$ is the data samples dimensionality, $m$ the random projections output dimensionality, and $p$ the number of random projections. Two types of dropouts can be introduced on the random projections leading to the encoding network training loss: a dropout on the projections themselves, and a dropout on the components of the projections. In the first case, the dropout removes entire projections, implying a selection, in accordance with the dropout rate, over the $p$-dimensional channel of the projecting random tensor. Components dropout implies a selection, with its own dropout rate, along the $d$-dimensional channel. The indexes selected for this dropout will then cancel the corresponding dimensions in the random projections, thus ignoring as many components among the inputs. The RPs are normalized again after the components dropout, when $m=1$, in accordance with \eqref{RPO_0}.

To respect the notation introduced, applications on images flatten the input pixels array into a $d$-dimensional vector before their projection. As an intuitive example, in the specific case $m=1$ which coincides with the original RPO, the random projections define a matrix $d \times p$: the projections dropout here would remove columns over the second dimension, whereas the components dropout would discard rows over the first dimension.

\subsection{Experiments}

	\begin{table}[H]
	\caption{Deep RPO-mean test AUCs with and without components and projections dropouts on Fashion-MNIST for 10 seeds (truncated mean AUC $\pm$ std). C. is components dropout rate, P. is projections dropout rate.}
	\label{dropouts-table}
	%\vskip 0.01in
	\begin{center}
	\begin{small}
	\begin{sc}
	\scalebox{1.0}{
	\begin{tabular}{lccccc}
	\toprule
	\# modes & 1 & 3 \\
	
	\midrule
	No dropout  & 89.00 $\pm$3.71& \textbf{78.71} $\pm$4.80\\
	C = 0.1		& \textbf{89.19} $\pm$3.57& 78.64 $\pm$4.85\\
	C = 0.3		& 89.18 $\pm$3.58& 78.64 $\pm$4.85\\
	C = 0.5		& \textbf{89.19} $\pm$3.57& 78.64 $\pm$4.85\\
	P = 0.1		& 89.05 $\pm$3.68& 78.51 $\pm$4.93\\
	P = 0.3		& 88.88 $\pm$3.80& 78.36 $\pm$4.97\\
	P = 0.5		& 88.67 $\pm$3.97& 78.43 $\pm$4.76\\
	\bottomrule	
	\end{tabular}
	}
	\end{sc}
	\end{small}
	\end{center}
	\vskip -0.1in
	\end{table}
	
Table \ref{dropouts-table} indicates there is no actual AUC increase when harnessing either of the dropouts put forward for the random projections leading to the outlyingness measure. Since no substantial performances improvement was reached using the dropouts individually, their combined effects were not studied.

\section{Potential for an extension to a semi-supervised setting}

\subsection{Extension description}

Semi-supervised anomaly detection goes beyond the scope of this work, but it is trivial to adapt deep RPO to SAD just like deep SVDD was transformed into deep SAD \cite{deepSAD}. Like deep SAD, one only needs to take into account the rare labeled anomalies responsible for the semi-supervision during training by inverting the distance to the normality location estimators in the training loss. In such a SAD context, the normalized distance $dist$ to the normality locations estimators will be counted as $\frac{1}{dist}$ if the sample considered is a labeled anomaly of the training set. Intuitively, this compels the neural network to project anomalies far from the locations estimators, while at the same time gathering normal samples around them. This reveals a danger for both unsupervised and semi-supervised anomaly detection in realistic settings: the contamination of the normal samples within the training set with unidentified and representative anomalies can make the neural network project anomalies close to locations estimators, rendering them hardly detectable. Besides, the ability of the network to generalize and efficiently reject anomalies far from normality reference points while concentrating normal samples around them implies a good representativity of both normal samples and anomalies in the training set. This hypothesis remains necessarily unmet since in most AD settings, anomalies are infinitely diverse.

\subsection{Experiments}

	\begin{table}[H]
	\caption{Semi-supervised anomaly detection with distance inversion as in deep SAD for deep RPO to take into account rare labeled anomalies during training. The SAD ratio denotes the percentage of the training set composed of labeled anomalies. Two anomalous classes are randomly picked for each seed to provide the labeled anomalies. Experiments conducted with either one or three modes in the normality on Fashion-MNIST for 10 seeds (truncated mean AUC $\pm$ std).}
	\label{SAD-table}
	%\vskip 0.01in
	\begin{center}
	\begin{small}
	\begin{sc}
	\scalebox{0.97}{
	\begin{tabular}{lc|cc}
	\toprule
	SAD method & SAD ratio & 1 & 3 \\
	
	\midrule
	deep SAD & 0.00 & 87.70 $\pm$5.30& 78.30 $\pm$5.02\\
	deep SAD & 0.01 & 88.08 $\pm$5.03& 83.49 $\pm$4.71\\
	deep SAD & 0.10 & 90.37 $\pm$4.00& 84.54 $\pm$4.87\\

	\midrule
	deep RP-SAD & 0.00 & 89.00 $\pm$3.71& 78.71 $\pm$4.80\\
	deep RP-SAD & 0.01 & 89.19 $\pm$3.60& 78.76 $\pm$4.90\\
	deep RP-SAD & 0.10 & 89.40 $\pm$3.46& 79.93 $\pm$5.30\\

	\bottomrule	
	\end{tabular}
	}
	\end{sc}
	\end{small}
	\end{center}
	\vskip -0.1in
	\end{table}

Deep SVDD, transformed into deep SAD, appears to more significantly benefit from the additional information provided by a small minority of labeled anomalies during the training. Nevertheless deep RPO also takes advantage of the latter to improve detection performances, confirming the generality of the distance inversion method to allow a location estimator based unsupervised AD to achieve SAD.

\section{Stability against affine transformation}

	\begin{table*}[t]
	\caption{Deep RPO-mean test AUCs with varying affine transformation coefficient $\alpha$ on Fashion-MNIST for 10 seeds (truncated mean AUC $\pm$ std). The AUC gap is the mean AUC error, computed over all seeds, with respect to the AUC obtained when $\alpha = 0$, i.e. the baseline case. In the first part of the table, $\alpha$ denotes the constant value along the affine transformation diagonal matrix. In the second part, the diagonal elements are randomly generated according to either a uniform or a gaussian standard distribution. No AUC gap is computed since the seed by seed comparison with the baseline AUC would be unfair, a mean AUC test being computed over 20 random picks of the diagonal matrix for each seed.}
	\label{affine-table}
	%\vskip 0.01in
	\begin{center}
	\begin{small}
	\begin{sc}
	%\scalebox{1.0}{
	\resizebox{0.8\textwidth}{!}{
	\begin{tabular}{lc|cc|ccc}
	\toprule
	AD method & $\alpha$ & 1 & AUC gap & 3 & AUC gap\\
	
	\midrule
	deep SVDD & 0.80 & 87.02 $\pm$5.56& -0.70 $\pm$1.05 & 76.49 $\pm$5.67 & -1.81 $\pm$0.86\\
	deep SVDD & 0.90 & 87.77 $\pm$5.24& +0.03 $\pm$0.29 & 78.02 $\pm$5.15 & -0.29 $\pm$0.29\\
	deep SVDD & 0.95 & 87.83 $\pm$5.22& +0.09 $\pm$0.14 & 78.31 $\pm$5.02 & -0.00 $\pm$0.16\\
	deep SVDD & 1.00 & 87.73 $\pm$5.24& $\pm$0.00 $\pm$0.00 & 78.31 $\pm$5.02 & $\pm$0.00 $\pm$0.00\\
	deep SVDD & 1.05 & 87.48 $\pm$5.30& -0.25 $\pm$0.20 & 78.05 $\pm$5.18 & -0.25 $\pm$0.28\\
	deep SVDD & 1.10 & 87.09 $\pm$5.40& -0.63 $\pm$0.50 & 77.55 $\pm$5.52 & -0.76 $\pm$0.75\\
	deep SVDD & 1.20 & 86.01 $\pm$5.71& -1.71 $\pm$1.32 & 75.98 $\pm$6.65 & -2.32 $\pm$2.13\\
	\midrule
	deep RPO & 0.80 & 88.53 $\pm$4.15& -0.72 $\pm$1.48 & 76.85 $\pm$5.28 & -1.77 $\pm$1.22\\
	deep RPO & 0.90 & 89.25 $\pm$3.65& -0.00 $\pm$0.59 & 78.33 $\pm$4.85 & -0.29 $\pm$0.57\\
	deep RPO & 0.95 & 89.33 $\pm$3.56& +0.07 $\pm$0.31 & 78.61 $\pm$4.79 & -0.01 $\pm$0.29\\
	deep RPO & 1.00 & 89.26 $\pm$3.54& $\pm$0.00 $\pm$0.00 & 78.63 $\pm$4.85 & $\pm$0.00 $\pm$0.00\\
	deep RPO & 1.05 & 89.01 $\pm$3.60& -0.24 $\pm$0.38 & 78.38 $\pm$5.05 & -0.24 $\pm$0.34\\
	deep RPO & 1.10 & 88.62 $\pm$3.76& -0.63 $\pm$0.84 & 77.91 $\pm$5.36 & -0.71 $\pm$0.74\\
	deep RPO & 1.20 & 87.48 $\pm$4.39& -1.77 $\pm$1.95 & 76.49 $\pm$6.25 & -2.13 $\pm$1.74\\
	\midrule
	deep SVDD & $U_{[0.9;1.1]}$ & 87.47 $\pm$5.10& & 78.15 $\pm$4.76 & \\
	deep SVDD & $U_{[0.8;1.2]}$ & 86.70 $\pm$5.52& & 77.52 $\pm$5.01 & \\
	deep SVDD & $U_{[0.6;1.4]}$ & 83.86 $\pm$7.21& & 75.67 $\pm$5.94 & \\
	deep SVDD & $U_{[0.5;1.5]}$ & 81.28 $\pm$8.74& & 73.64 $\pm$7.10 & \\
	deep SVDD & $N(0,1)$		 & 52.20 $\pm$16.01& & 50.56 $\pm$12.37 & \\
	\midrule
	deep RPO & $U_{[0.9;1.1]}$ & 88.76 $\pm$3.61&  & 78.54 $\pm$4.70 & \\
	deep RPO & $U_{[0.8;1.2]}$ & 87.99 $\pm$3.99&  & 77.97 $\pm$5.03 & \\
	deep RPO & $U_{[0.6;1.4]}$ & 85.12 $\pm$5.73&  & 76.32 $\pm$6.05 & \\
	deep RPO & $U_{[0.5;1.5]}$ & 82.49 $\pm$7.35&  & 74.59 $\pm$7.19 & \\
	deep RPO & $N(0,1)$		 & 52.02 $\pm$16.16&  & 50.05 $\pm$10.77 & \\
	
	\bottomrule	
	\end{tabular}
	}
	\end{sc}
	\end{small}
	\end{center}
	\vskip -0.1in
	\end{table*}
	
An affine transformation, defined as a constant multiplication of every component of the input representation of the samples, is applied to challenge the affine stability of the AD methods performances once the training is over. This affine transformation, defined by the constant $\alpha$ shown in the upper part of Table \ref{affine-table}, breaks the normalization of the inputs features before their presentation to the neural network first layer. The experiments results suggest that deep RPO and deep SVDD are comparably stable with respect to the input transformation considered, and that such transformation does not trigger a drop in AUC. In addition, one can notice that the average test AUC slightly increases in some cases with the affine data disturbance. The lower part of Table \ref{affine-table} reports the results where instead of a constant diagonal matrix applying $\alpha$ to each input component, another diagonal matrix is used for which the diagonal coefficients are generated using either a random uniform or a standard gaussian distribution. Again, deep RPO and deep SVDD show comparable stability when confronted with the more distorting affine transformations. Looking at the standard deviations overall, deep RPO seems slightly more stable.

\section{Additional experiments on tabular data}

	\begin{table}[H]
	\caption{Deep RPO-Mean, RPO-Max and the baseline deep SVDD on the satellite dataset for 20 seeds (truncated mean AUC $\pm$ std).}
	\label{satellite}
	%\vskip 0.01in
	\begin{center}
	\begin{small}
	\begin{sc}
	\scalebox{1.0}{
	\begin{tabular}{lc}
	\toprule
	Method & mean test AUC $\pm$ std \\
	
	\midrule
	Deep SVDD	& 68.23 $\pm$5.53\\
	RPO-Max	& 64.89 $\pm$2.67\\
	Deep RPO-Mean	& \textbf{73.01} $\pm$5.93\\
	\bottomrule	
	\end{tabular}
	}
	\end{sc}
	\end{small}
	\end{center}
	\vskip -0.1in
	\end{table}
	
Since AD on MNIST, Fashion MNIST and CIFAR10 is very common and excellent performances have already been obtained on these datasets using self-supervised learning, we compare the highlighted shallow and deep methods of our main results in Table \ref{main-table} on less common tabular data. As can be seen in Table \ref{satellite}, Deep SVDD remains our baseline. A satellite dataset is chosen \footnote{\url{http://odds.cs.stonybrook.edu/satellite-dataset/}}. The data stems from the original Statlog (Landsat Satellite) dataset from UCI machine learning repository \footnote{\url{https://archive.ics.uci.edu/ml/datasets/Statlog+\%28Landsat+Satellite\%29}}, where the smallest three classes are combined to form the outlier class, while the other classes define the inlier class. As for the previous experiments, deep SVDD and deep RPO-Mean share the same neural network architecture and training hyperparameters, to produce a fair comparison. The improvement provided by Deep RPO-Mean is confirmed. The number of RPs used in the latent RPO was set to 500, since the output dimensionality of 8 of the neural network is significantly lower. This in turn is due to the low input data dimensionality for the neural network, the input samples being 1D vectors defined by 36 values. The neural networks were always trained for 80 epochs, and the test AUC retained as the model performance for each seed is the one associated with the best epoch with respect to the validation set AUC. The results also put forward the contribution of the trainable neural network projecting data samples, deep SVDD performing better then the shallow method RPO-Max. Finally, the standard deviation of the performances appears to be higher for deep methods.

\end{document}